\documentclass[prl,aps,twocolumn,floatfix,10pt]{revtex4-1}

\usepackage{graphicx} 
\usepackage{amsmath}
\usepackage{hyperref}

\begin{document} 
\title{Non-perturbative method to compute thermal correlations in one-dimensional systems: A brief overview}

\author{Stefan Beck$^{1,2}$, Igor E. Mazets$^{1,2}$, and Thomas Schweigler$^1$} 
\affiliation{
$^1$\, Vienna Center for Quantum Science and Technology, Atominstitut, TU~Wien,~Stadionallee~2,~1020~Vienna,~Austria \\
$^2$\, Wolfgang Pauli Institute c/o Fakult\"{a}t f\"{u}r Mathematik,
Universit\"{a}t Wien, Oskar-Morgenstern-Platz 1, 1090 Vienna, Austria}

\begin{abstract} 
We develop a highly efficient method to numerically simulate thermal fluctuations and correlations in non-relativistic continuous bosonic
one-dimensional systems. 
We start by noticing the equivalence of their description through the transfer matrix formalism and a Fokker-Planck equation for a distribution evolving in space.
The corresponding stochastic differential (It\={o}) equation is very suitable for computer simulations, allowing the calculation of arbitrary correlation functions. 
As an illustration, we apply our method to the case of two tunnel-coupled quasicondensates of bosonic atoms.
\end{abstract}
\maketitle

One-dimensional (1D) systems attract much attention because their dynamics is strongly affected by the restricted phase space available for scattering~\cite{Giamarchi04,Cazalilla11}. Experimentally available 1D systems range from ultracold atomic gases~\cite{bloch2008many,proukakis2017universal} to slow-light polaritons~\cite{chang2008} as well as superfluid $^4$He atoms adsorbed in nanometer-wide pores~\cite{Wada01}. They allow the implementation of fundamental theorectical models such as the celebrated Lieb-Liniger model~\cite{Lieb63,Lieb63b} as well as the quantum Luttinger-liquid model~\cite{Luttinger63}. Several recent experimental studies underline the importance of 1D systems as a testbed for theoretical ideas, in and out of equilibrium~\cite{Langen2015,Bordia2017,Rauer17,Schweigler17}.

The rapid progress in the preparation, manipulation and characterization of experimental systems, especially in the realm of ultracold atoms, also leads to a need for ever improving theoretical descriptions. In particular, the recent measurement of higher-order correlation functions in 1D quasi-condensates~\cite{Schweigler17} calls for novel theoretical methods beyond the perturbative approach.

In this Letter, we report the development of a universal method to calculate thermal correlations of multicomponent bosonic fields in the mean-field approximation in 1D, which is a generalization of our early method  for Gaussian fluctuations by means of the Ornstein-Uhlenbeck stochastic process~\cite{stimming2010fluctuations}. 
Our new method is applicable to a wide variety of non-relativistic continuous 1D bosonic systems with local interactions. It provides highly efficient numerical sampling of classical fields with the statistics given by the thermal equilibrium. In this Letter, we present the method and apply it to the case of two tunnel-coupled 1D quasicondensates~\cite{Goldstein97,Whitlock03}. A more extensive presentation including a detailed derivation of the method can be found in~\cite{we-PRA}.   

We consider a 1D complex field (a mean-field approximation for a quantum many-body problem) with ${\cal M}/2$ 
components $\psi _j$ (${\cal M}$ is an even integer number) or, equivalently, with ${\cal M}$ real components 
\begin{equation} 
q_{2j-1}=\mathrm{Re}\, \psi _j, \quad  
q_{2j} = \mathrm{Im}\, \psi _j, \quad j=1,\, 2, \, \dots \, ,\, {\cal M}/2.
\label{qj}
\end{equation} 
Without loss of generality, we assume that all the components are characterized by the same mass $m$. The Hamiltonian 
function of the system is then
\begin{eqnarray} 
H&=&\int _{-L/2}^{L/2}
dz\, \Bigg{\{ }\sum _{j=1}^{\cal M} \left[ \frac {\hbar ^2}{2m} \left( \frac {\partial q_j}{\partial z}\right)^2-\mu _j q_j^2
\right]  \nonumber \\ && +  V (q_1, \, \dots \, ,\, q_{\cal M}) \Bigg{ \} }.
\label{hm1}
\end{eqnarray} 
Here $\mu _j$ is the chemical potential for the $j$th component. 
Since $q_{2j-1}$ and $q_{2j}$ are the real and imaginary part of the same complex field, there are only ${\cal M}/2$ independent chemical potentials. 
$V (q_1, \, \dots \, ,\, q_{\cal M})$ represents the local interaction energy density, which does not explicitly depend on $z$ (homogeneous system).
We consider the thermodynamic limit $L\rightarrow \infty $, while the mass density remains constant. 

We are interested in equal-time  correlations of the observables ${\cal F}^{(i)}\vert _{z_i}=
{\cal F}^{(i)} [q_1(z_i), \, \dots \, ,\, q_{\cal M}(z_i)]$ 
measured at different points $z_i$, $i=1,\, 2, \, \dots \, ,\, l$ at  
equilibrium with temperature $T$. 
Generally all the chosen observables ${\cal F}^{(i)}$ may be different.
Starting from the transfer matrix formalism~\cite{Scalapino72,Krumhansl75,Currie80} we can write the correlations as  
\begin{align} 
&
\langle {\cal F}^{(1)}\vert _{z_1} {\cal F}^{(2)}\vert _{z_2} \, \dots \, 
{\cal F}^{(l)}\vert_{z_l}\rangle = 
\sum _{\nu _1, \dots ,\nu_{l-1} } \langle 0|{\cal F}^{(l)}|\nu_{l-1}\rangle \dots  
~ \nonumber \\ & \quad \times 
\langle \nu_2|{\cal F}^{(2)}|\nu_{1}\rangle  \langle \nu_1|{\cal F}^{(1)}|{0}\rangle 
\prod _{i=1}^{l-1}e^{-(\kappa _{\nu_i} -\kappa _0)(z_{i+1}-z_i)} .
 \label{FFF} \end{align} 
Here the spatial points are ordered as $z_l>\dots >z_2>z_1$.  
The $\kappa_\nu$ are eigenvalues, and $|\nu \rangle $ are normalized eigenfunctions of the auxiliary Hermitian  operator~\cite{Scalapino72}
\begin{equation} 
\hat K=\sum _{j=1}^{\cal M} \left( -D\frac {\partial ^2}{\partial q_j^2}-\frac {\mu _j}{k_BT}q_j^2\right)  +
\frac {V (q_1, \, \dots \, ,\, q_{\cal M})}{k_BT} 
\label{K} 
\end{equation}
where
\begin{equation}
D=mk_BT/(2\hbar ^2).
\end{equation}
We assume that the  lowest eigenvalue denoted by $\kappa _0$ is non-degenerate and denote the corresponding 
eigenfunction (the ``ground state") 
by  $|0\rangle$.

$\hat K $ resembles a Hamiltonian for a single quantum particle in an ${\cal M}$-dimensional space. 
Note that the dimension of the eigenvalues is inverse length, not energy. We assume that the interaction 
$V\rightarrow +\infty $ for $q_j\rightarrow \pm \infty $, which makes the system stable~\cite{Scalapino72}. 
Therefore the boundary conditions to the equation $\hat K |\nu \rangle = \kappa_\nu |\nu \rangle $ 
require the eigenfunctions to vanish if $q_j\rightarrow \pm \infty $. 
The resulting functions  $|\nu \rangle $ form a complete, orthonormal basis. 

 For $\langle \nu^\prime |{\cal F}|\nu \rangle $ the standard quantum-mechanical definition of a matrix element applies. 
 Setting $l = 1$ and using ${\cal F}^{(1)}\vert _{z_1} = \prod _{j=1}^{\cal M}\delta (q_j-q_j^\prime )$ in Eq.~(\ref{FFF}) we can see that the thermal equilibrium distribution of the local 
values of the fields $q_j^\prime $ is given by
\begin{align}
\begin{split}
&W_{\mathrm{eq}}
(q_1^\prime , \, \dots \, ,\, q_{\cal M}^\prime ) = \langle 0|{\cal F}^{(1)}|0\rangle \\ &= \langle 0|\prod _{j=1}^{\cal M}\delta (q_j-q_j^\prime )|0\rangle =
\left| \langle q_1^\prime , \, \dots \, ,\, q_{\cal M}^\prime |0\rangle \right| ^2 . \label{eq:W0}
\end{split}
\end{align} In what follows, we denote the 
ground-state eigenfunction by $\Psi _0(q_1,\, \dots \, ,\, q_{\cal M})\equiv \langle q_1, \, \dots \, ,\, q_{\cal M}|0\rangle $.

Calculating correlation functions directly from Eq.~(\ref{FFF}) might be a challenging task, because of the need to know many eigenvalues and eigenvectors of $\hat K $. However, one can show that Eq.~(\ref{FFF}), which represents the most general classical $l$-point correlation function, is equivalent to the result following from the joint probability density for a stationary stochastic process 
being described by the Fokker-Planck equation~\cite{Risken89,we-PRA} 
\begin{eqnarray} 
&& \frac {\partial W(q_1,\, \dots \, ,\, q_{\cal M})}{\partial z}= \sum _{j=1}^{\cal M} \left \{ 
D \frac {\partial ^2}{\partial q_j^2} W(q_1,\, \dots \, ,\, q_{\cal M}) \right. 
\nonumber \\ && \quad 
-\left.  \frac \partial {\partial q_j}[ A_{q_j}(q_1,\, \dots \, ,\, q_{\cal M})W(q_1,\, \dots \, ,\, q_{\cal M})]\right \} 
\label{FPE}
\end{eqnarray} 
with the drift vector
\begin{eqnarray} 
A_{q_j}&=&D \frac {\partial \ln W_\mathrm{eq} (q_1,\, \dots \, ,\, q_{\cal M})}{\partial q_j} \nonumber \\
&=&2D \frac {\partial \ln |\Psi _0 (q_1,\, \dots \, ,\, q_{\cal M})|}{\partial q_j}. 
\label{A} 
\end{eqnarray} 
Here the 1D coordinate $z$ has the role usually played by the time. 
$W_\mathrm{eq}(q_1,\, \dots \, ,\, q_{\cal M})$ defined in Eq.~\eqref{eq:W0} is the stationary solution of Eq.~(\ref{FPE}) with $A_{q_j}$ defined by Eq.~(\ref{A}). 
Note, that $\Psi _0$ possesses all the properties of a ground-state function of a Hamiltonian problem, in particular, 
for all finite $q_j$'s it is non-zero and, hence, $A_{q_j}$ has no singularities. 

Direct numerical integration of the Fokker-Planck equation in a multidimensional space is a challenging task. We 
recall instead the equivalence of the Fokker-Planck equation and the stochastic differential It\={o} equation~\cite{Risken89,GardinerBook} 
\begin{equation} 
dq_j = A_{q_j}dz + \sqrt{2D}\, dX_j,
\label{Ito} 
\end{equation} 
where $dX_j$ are infinitesimally small, mutually uncorrelated, random terms obeying
Gaussian statistics with zero mean and the variance
equal to $dz$: $\overline{dX_j}=0$, $\overline{dX_j dX_{j^\prime }}=\delta _{jj^\prime }dz$. Here the bar denotes averaging over the ensemble of realizations of the 
stochastic process. The initial values (say, at $z=0$) of the fields for each realization are obtained by (pseudo)random 
sampling their equilibrium distribution $W_\mathrm{eq}$. The subsequent numerical integration of Eq.~(\ref{Ito}) and 
averaging over many realizations yields the correlation functions. 

Eq.~(\ref{Ito}) is therefore  a generalization of our previous method  to simulate  
the classical thermal fluctuations in a system described by a quadratic Hamiltonian using the Ornstein-Uhlenbeck 
stochastic process~\cite{stimming2010fluctuations}
to the case of the arbitrary local interaction $V$. The main advantage of the stochastic \textit{ordinary} differential Eq.~\eqref{Ito} 
is that its numerical integration is much simpler and less resource-consuming than the integration of the 
\textit{partial} differential   Eq.~(\ref{FPE}) on a multidimensional grid. 
The main computational difficulty is now reduced  to the precise determination 
of $\Psi _0$. 
The determination of 
all other eigenfunctions and eigenvalues 
appearing in Eq.~(\ref{FFF}) is actually not necessary.

We apply our method to the calculation of thermal phase and density fluctuations of two tunnel-coupled 1D quasicondensates of 
ultracold bosonic atoms. The quasicondensates in the right (\textit{R}) or in the left (\textit{L}) 
1D atomic waveguide are described in 
the mean-field approximation by complex classical fields $\psi _R\equiv q_1+iq_2$ and $\psi _L=q_3+iq_4$, respectively.
Alternatively, it is possible to express these complex fields $\psi _\varsigma =\sqrt{n_\varsigma }e^{i\theta _\varsigma }$, 
$\varsigma =R,\, L$, through the quasicondensate atom-number densities $n_{R,L} $ and phases $\theta _{R,L} $~\cite{Mora03}. 
This system is described by Eq.~(\ref{hm1}) with the interaction term~\cite{Goldstein97,Whitlock03} 
\begin{align} 
V(q_1, q_2, q_3, q_4) = &\frac g2 \left[ \left(q_1^2+q_2^2\right)^2+\left(q_3^2+q_4^2\right)^2 \right] \nonumber \\ &-2\hbar J \left(q_1 q_3 + q_2 q_4\right), 
\label{VgJ}
\end{align}
where $g$ is the strength of the contact interaction of atoms in 1D and $J$ is the single particle tunneling rate.
The tunneling provides exchange of atoms between the two waveguides, therefore atoms in both of them have the same 
chemical potential $\mu _R=\mu _L \equiv \mu $. The mean 1D atom-number density in each of the waveguides that 
corresponds to this chemical potential is denoted by $n_\mathrm{1D}$.  

It is convenient to parametrize the density variables as $n_\varsigma = r^2_\varsigma n_\mathrm{1D}$. Then we obtain 
\begin{equation}
\hat{K} =\frac 1{\lambda_T }\left[ \hat{\cal K}_R + \hat{\cal K}_L - 2 b \ r_R r_L \cos(\theta_R-\theta_L)\right] , \label{K_dens_phase}
\end{equation}
where
\begin{align}
\begin{split}
\hat{\cal K}_{\varsigma} = &-\left( \frac{\partial^2}{\partial r_\varsigma^2} + 
\frac{1}{r_\varsigma} \frac{\partial}{\partial r_\varsigma} + \frac{1}{r_\varsigma^2} \frac{\partial^2}{\partial \theta_\varsigma ^2} \right)  
\\ &+ \alpha \ r_\varsigma^2 \left[ r_\varsigma^2 - 2 \left( \tilde{\mu} - \frac{b}{2 \alpha} \right) \right] .
\label{KKo} 
\end{split}
\end{align}
The dimensionless parameters of the problem are 
\begin{equation}
\alpha = 
\frac {\lambda _T^2}{4\xi_\mathrm{h} ^2} , 
\qquad 
b =\frac { \lambda_T^2}{8 l_J^2},      \label{def.beta} 
\end{equation}
where $\xi_\mathrm{h} =\hbar /\sqrt{gn_\mathrm{1D} m}$ is the quasicondensate healing length,  
$\lambda_T = 2 \hbar^2 n_\mathrm{1D}/(m k_B T)$ is the thermal coherence length and $l_J = \sqrt{\hbar/(4 m J)}$ 
is the typical length of the relative phase locking~\cite{stimming2010fluctuations,grivsins2013coherence}. 
The parameters $\alpha$ and $b$ can be understood as the ratio of the energies of the mean-field repulsion 
and of the tunnel coupling, respectively, to the kinetic energy of an atom localized at the length scale of the order of $\lambda _T$. 
Note that 
$\tilde \mu =\mu/(gn_\mathrm{1D} )$ is not a free parameter, but 
has to be chosen such that the average 1D density equals $n_\mathrm{1D}$ 
in both waveguides, i.e., $\langle r_L^2\rangle = \langle r_R^2\rangle =1$.
Therefore the eigenstates of $\hat{K}$  depend on $\alpha$ and $b$ only. 
Since $D = n_\mathrm{1D}/\lambda_T$, the solution of the It\={o} equation \eqref{Ito} also depends on the scaled distance $z/\lambda_T$.

Finding the ground state of the operator $\hat K$ is still a formidable task. However, the general structure of the operator \eqref{K} [and, hence, of Eq.~\eqref{K_dens_phase}]  that 
contains only local pairwise interactions admits for a solution. First of all, we notice that Eq.~\eqref{K_dens_phase} is invariant with respect to 
simultaneously shifting  both the angles $\theta_R $ and $\theta_L$ by the same value. The (non-degenerate) ground state 
$\Psi _0$ must be independent of $\theta _+=(\theta_R + \theta_L)$. 
Using a proper basis for the expansion in $r_R$, $r_L$ and $\theta_-=(\theta_R - \theta_L)$ we are able to obtain $\Psi _0(r_R,r_L,\theta _-)$, for details see~\cite{we-PRA}.
After having obtained the ground state, we can integrate Eq.~(\ref{Ito}) by the forward Euler method, using a pseudo-random generator to simulate the random term.

In the following we will focus on discussing the relative phase fluctuations $\theta_-$, because they can be accessed experimentally through matter-wave interferometry~\cite{Schumm05}.
While it only makes sense to discuss the phase $\theta_-(z)$ modulo $2\pi$ for a single point, the unbound phase differences $\theta_-(z) - \theta_-(z')$ between two different points $z$ and $z'$ have a physical meaning. We obtain continuous phase profiles from the numerical samples of $\psi_{R,L}$ through phase unwrapping, i.e., by assuming that $\theta_-$ between neighbouring points on the numerical grid does not differ by more then $\pi$. The same procedure has been applied to experimental data in Ref.~\cite{Schweigler17}. 

We compare the results for the two coupled quasicondensates  to the predictions of the sine-Gordon (SG) model  
\begin{align} 
\begin{split}
\label{eq:SG}
{H}_{\mathrm{SG}} =   &\int _{-L/2}^{L/2}{d}z \left[ gn_-^2 + \frac{\hbar^2 n_\mathrm{1D}}{4m} 
\left(\frac {\partial {\theta_-}}{\partial z}\right) ^2 \right] \\
&- \int _{-L/2}^{L/2}{ {d}z ~ 2 \hbar J n_\mathrm{1D} \cos{\theta_-} } \, \mathrm{,}
\end{split}
\end{align} 
where $n_-=(n_R-n_L)/2$. 
The SG model was proposed as an effective model for the coupled quasicondensates~\cite{Gritsev07}. 
Its validity in a certain parameter regime was recently confirmed experimentally~\cite{Schweigler17}.  
Due to the simpler nature of the model we can  obtain results directly from the transfer matrix formalism~\cite{we-PRA}, without 
numerical implementation of Eq.~(\ref{Ito}). 
Since Eq.~\eqref{eq:SG} does not contain terms coupling $n_-$ to $\theta _-$, the relative density fluctuations can be integrated out 
and  the relative phase correlations are fully determined by the eigensystem of  
the auxiliary Hermitian operator for the  SG model~\cite{grivsins2013coherence}  
\begin{equation}
\hat{K}^{\mathrm{SG}} = \frac 1{\lambda_T }\left( - 2 \frac{\partial^2}{\partial \theta_-^2} - 2 b  \cos\theta_-\right) ,
\label{KSG} 
\end{equation}
which can be formally obtained from Eq.~\eqref{K_dens_phase} by setting $r_L = r_R \equiv 1$, $\partial /\partial r_{R,L}\equiv 0$, and 
$\partial ^2/\partial \theta _{R,L}^2 \equiv \partial ^2/\partial \theta _-^2$. Note that Eq.~(\ref{KSG}) does not contain the 
parameter $\alpha $, i.e., the equal-time phase correlations in the SG model at finite temperature do not depend on the 
atomic interaction strength.

\begin{figure}[h]
	\centering
	\includegraphics[width=\linewidth]{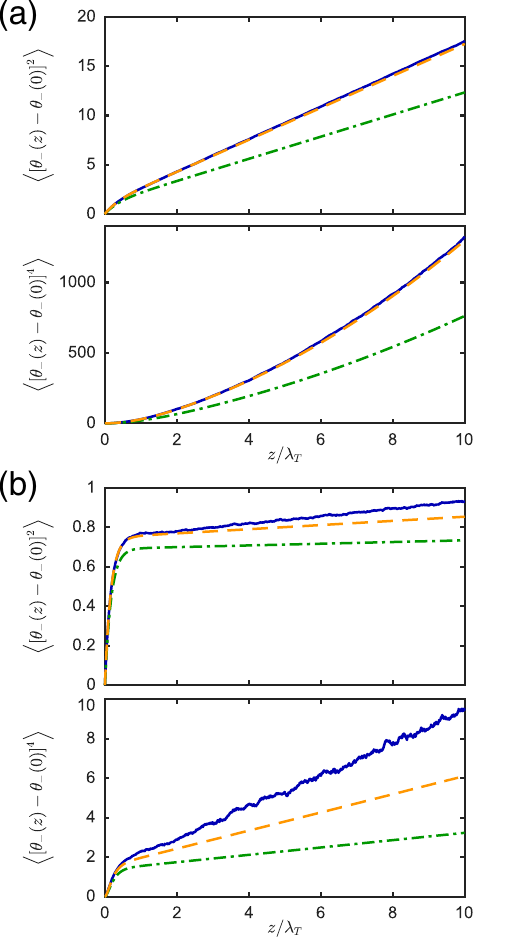}
	\caption{(Color online.) Second and fourth moment of the relative phase difference between two points along the 1D direction $z$. Results for $\alpha = 100$ and {(a)} $b = 1$, {(b)} $b = 5$. {(a)} In the intermediate phase-locking regime ($\left\langle \cos(\theta_-) \right\rangle = 0.58 $) we  observe good agreement between the two coupled 1D quasicondensates (solid blue lines) and the sine-Gordon model with the rescaled parameters (dashed orange lines). Clear deviations from the  the sine-Gordon model without rescaling of the parameters (green dash-dotted lines) are visible. {(b)} For strong phase locking ($\left\langle \cos(\theta_-) \right\rangle = 0.83 $) we get clear deviations also for the rescaled sine-Gordon theory.}
	\label{fig:fig1}
\end{figure}

For small and intermediate phase-locking the results for the full model agree with the predictions of the SG model with the rescaled parameters $\tilde{\lambda}_T =\lambda_T / \langle 1/r_\varsigma^2 \rangle _\mathrm{reg} $ and $\tilde{b} =b \ \langle r_R r_L \rangle / \langle 1/r_\varsigma^2 \rangle _\mathrm{reg} $~\cite{we-PRA}. Here $\langle 1/r_\varsigma^2 \rangle _\mathrm{reg}$ 
represents the regularized mean inverse density (in dimensionless units), for symmetry reasons the expectation value is the same for $\varsigma = L, R$. Without regularization, $\langle 1/r_\varsigma^2 \rangle $ diverges 
logarithmically. Different ways to regularize have been tested and all yielded very close results. 
Fig.~\ref{fig:fig1}(a) shows the results for $b = 1$, which corresponds to intermediate phase locking. One sees good agreement between the results for the full calculation and the rescaled SG model. For stronger phase-locking $b = 5$  deviations are clearly visible [Fig.~\ref{fig:fig1}(b)].

\begin{figure}
	\centering
	\includegraphics[width=\linewidth]{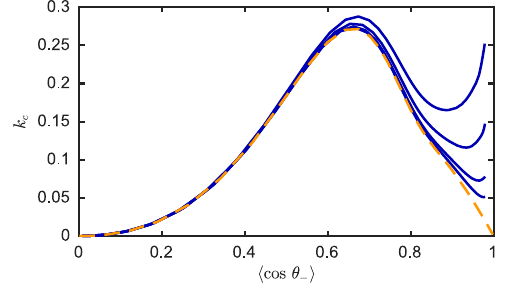}
	\caption{(Color online.) Circular kurtosis $k_c$ as defined in Eq.~\eqref{circ_kurt}. The solid blue lines represent the results for the coupled quasicondensates, the different lines represent, from top to bottom, $\alpha = 100, 200, 500, 1000$. The dashed orange line represents the sine-Gordon prediction.}
	\label{fig:fig2}
\end{figure}

Note that it is not possible to achieve agreement by using a different rescaling of parameters in the strong-coupling case. One can best see this from single-point expectation values calculated from $W_{\mathrm{eq}}$ \eqref{eq:W0} . They only depend on $b$ for the SG model and on $\alpha$ and $b$ for the coupled quasicondensates.
We analyze  the circular kurtosis~\cite{fisher93}
\begin{equation}
k_c = \frac{\left\langle \cos(2 \theta_-) \right\rangle - \left\langle \cos  \theta_- \right\rangle^4}{(1 - \left\langle \cos \theta_- \right\rangle)^2}, \label{circ_kurt}
\end{equation}
which is a measure for the non-Gaussianity of the underlying distribution of $\theta_-$.
Fig.~\ref{fig:fig2} shows $k_c$  as a function of $\left\langle \cos \theta_- \right\rangle$. One can see the deviation of the exact results from the predictions of the SG model for $\langle \cos \theta _-\rangle \approx 1$.

The deviation from the SG theory is bigger for smaller values of $\alpha$, i.e. for higher temperatures 
or lower densities. The density fluctuations are the physical reason. The higher $\alpha $ (i.e., the more pronounced the 
effect of interatomic repulsion), the more suppressed are the density fluctuations. The accuracy of the 
SG description is thus increased. The good agreement of the experimental data of Ref.~\cite{Schweigler17} with the SG model can be explained 
by rather a high value $\alpha \approx 600$. For $\alpha = 100$ the discrepancy between the full description of two 
coupled quasicondensates and the SG model becomes well pronounced. 
However, experimental measurements in this parameter regime are challenging due to the finite resolution of the imaging system.  

Note that the non-Gaussianity for intermediate phase-locking (intermediate values of  $\left\langle \cos \theta_- \right\rangle$) and strong phase-locking ($\left\langle \cos \theta_- \right\rangle \approx  1$) has different physical origins. For intermediate phase-locking, $W_\mathrm{eq}(r_R,r_L,\theta_-)$ as a function of $\theta _-$ for fixed $r_R, r_L$ is non-Gaussian in the relevant range of $r_R$ and $r_L$ (close to 1). For strong phase-locking this is not the case any more. The distribution of $\theta_-$ for different points $r_R$, $r_L$ is approximately Gaussian, with the variance depending on $r_R$, $r_L$. Therefore, averaging over different points leads to an overall distribution for $\theta _-$ which is non-Gaussian.

To conclude, we have developed a versatile method for calculating thermal expectation values for 1D systems. We applied the method to the case of two tunnel-coupled 1D quasicondensates. We identified the cases when this system can be described by the simpler sine-Gordon model and when this description breaks down. 

Our non-perturbative method is applicable to basically all stable continuous 1D bosonic systems with local interactions as long as thermal fluctuations describable by classical fields dominate. 
Additional requirements are that the system is homogeneous and non-relativistic.
The main advantage of the presented method is its computational efficiency. Calculating the $1.2 \times 10^5$ realizations used for Fig.~\ref{fig:fig1} takes around 2 hours on a desktop computer, which is at least by an order of magnitude shorter  than what more traditional methods like stochastic Gross-Pitaevskii (SGPE)~\cite{Blakie08} would need. 
Moreover, we should mention the robustness of our method in the presence of (quasi)topological excitation. Such excitations often comprise a problem when using methods based on the evolution in presence of a noise term (SGPE) or some sort of Metropolis-Hastings algorithm ~\cite{Grisins13}. We therefore believe that our method will find its application in a broad research area.

The authors thank S. Erne, V. Kasper, and J. Schmiedmayer for helpful discussions. We acknowledge financial
support by the by the Wiener Wissenschafts und Technologie Fonds (WWTF) 	
via the grant MA16-066  
and by the EU via the ERC advanced grant QuantumRelax (GA 320975). This work was also supported by the Austrian Science Fund (FWF) via
the project P~25329-N27 (S.B., I.M.), the SFB ISOQUANT No. I 3010-N27, and the Doctoral Programmes W~1245-N25 ``Dissipation und Dispersion in nichtlinearen partiellen Differentialgleichungen" 
(S.B.) and W~1210-N25 CoQuS (T.S.).

\bibliography{biblioA1}

\end{document}